\documentclass[conference]{IEEEtran}
\IEEEoverridecommandlockouts
\usepackage{cite}
\usepackage{amsmath,amssymb,amsfonts}
\usepackage{algorithmic}
\usepackage{graphicx}
\usepackage{textcomp}
\usepackage{float}
\usepackage{mathtools}

\usepackage{graphicx}
\graphicspath{ {./} }
\usepackage{xcolor}

\begin{document}
	
\title{Protecting IoT Servers Against Flood Attacks with the Quasi Deterministic Transmission Policy}
	
	\author{\IEEEauthorblockN{Erol Gelenbe}
		\IEEEauthorblockA{\textit{Institute of Theoretical \& Applied Informatics}\\
			Polish Academy of Science, 44-100 Gliwice, PL\\
			\& CNRS I3S Lab., Universit\'{e} C\^{o}te d'Azur, FR\\
			\& Ya\c{s}ar University, Bornova, Izmir, TR\\
			ORCID:0000-0001-9688-2201}
		 \and
		
		\IEEEauthorblockN{Mohammed Nasereddin}
		\IEEEauthorblockA{\textit{Institute of Theoretical \& Applied Informatics}\\
			Polish Academy of Science, 44-100 Gliwice, PL\\
		ORCID: 0000-0002-3740-9518}}
	
	\maketitle

	\begin{abstract}
		 IoT Servers that receive and process packets from IoT devices should meet the  QoS needs of incoming packets, and support Attack Detection software  that analyzes the incoming traffic to identify and discard packets that may be part of a Cyberattack. Since UDP Flood Attacks can overwhelm IoT Servers by creating congestion that paralyzes their operation and limits their ability to conduct timely Attack Detection, this paper proposes and evaluates a simple architecture to protect a Server that is connected to a Local Area Network, using a Quasi-Deterministic Transmission Policy Forwarder (SQF) at its input port. This  Forwarder shapes the incoming traffic, sends it to the Server  in a manner which does not modify the overall delay of the packets, and avoids congestion inside the Server. The relevant theoretical background is briefly reviewed, and  measurements during a UDP Flood Attack are provided to compare the Server performance, with and without the Forwarder. It is seen that during a UDP Flood Attack, the Forwarder protects the Server from congestion allowing it to effectively identify Attack Packets. On the other hand, the resulting Forwarder congestion can also be eliminated at the Forwarder with ``drop'' commands generated by the Forwarder itself, or sent by the Server to the Forwarder.
		\end{abstract}
	
	\begin{IEEEkeywords}
	Internet of Things (IoT), Cyberattack Detection, Congestion, Traffic Shaping, Quasi-Deterministic Transmission Policy (QDTP), Quality of Service
\end{IEEEkeywords}







  \section{Introduction}

With some $30$ Billion devices on the Internet \cite{Cisco2020},
many types of anomalies have been observed as a result of cyberattacks
 \cite{zanella2013m2m,jin2017recursive, liu2017novel, tello2018performance}, including Denial of service (DoS) attacks that disable target systems by flooding them with huge streams of requests \cite{Cyber23}. While many such attacks go unreported
 when they occur, just one Distributed DoS attack in $2017$ targeting Google, 
 compromised $180,000$ web servers which flooded Google servers at overall
 bitrates of $2.54$ Tera-bits/sec \cite{Cloudflare1}. Other attacks aim mainly at the IoT  \cite{Mahmoud,Farooq,hp,Spilios}, while Botnet attacks \cite{Statt} are particularly vicious since they spread by inducing their victims to become attackers \cite{Tushir_impactsOfMirai,Sinanovic,Manos}.
 
 UDP Flood attacks \cite{Cloudflare2} are also exploited by Botnets to create massive congestion that overwhelms network nodes and ports. Using spoofed-source-address UDP packets, they cause their victims to crash due to high traffic volumes, 
 creating denial of service,  causing lost data and resulting in missing and incomplete readings 
 of the data carried by legitimate IoT traffic \cite{Iqbal,Chin}.

 \subsection{Prior Work}
 Because of the concern about cybersecurity, there is a large literature on cyberattacks and Attack Detection (AD) methods 
 \cite{Douligeris,Cyber22,Cyber23}. These methods are typically evaluated 
for accuracy  using statistical methods \cite{Spilios,AHMETOGLU22}, and various Machine Learning based AD algorithms are often tested under ideal conditions on general purpose
 computers \cite{Banerjee,Tuan,Al-Issa,Guven,CDIS}, where attack traffic is treated as data, and the attack's overload on the processing capacity and performance of the victim node is not taken into account. 
 
 Various AD test-beds \cite{mirkovic2009test} for cyber-physical 
 and IoT networks are presented in \cite{Kaouk2018,annor2018development,waraga2020design}. Experiments on windfarms under SYN attacks are discussed in \cite{singh2020testbed} and other experimental IoT security studies can be found in \cite{ghaleb2016scada,tesfahun2016scada,reutimann2022simulating}.
 Data collection and display for  flood attacks are discussed in \cite{park2018test} while in \cite{arthi2021design} real-time data collection for IoT DNS attacks is presented. Denial of Service (DoS) attacks against Software Defined Networks (SDN) that support the IoT have also been studied \cite{wright2019testbed}.  However this prior work relates to attack emulation environments that do not include the overload caused by attacks, as recently discussed for 
 autonomous vehicles\cite{sontakke2022impact} and IoT servers \cite{NASEREDDIN2023}.
 
 \subsection{Motivation and Research Plan}
 
 Thus the present paper is motivated by the need to:
 \begin{itemize}
 \item  Experimentally evaluate the effect of IoT Server overload during an ongoing UDP Flood attack, 
 \item Understand the attack's impact on the Server's capacity to carry out Attack Detection (AD) and other useful processing functions, 
 \item Demonstrate a system architecture, and a traffic shaping policy  \cite{ICC22} that was initially proposed to mitigate the IoT's Massive Access Problem (MAP) \cite{Rodoplu2019,Rodoplu2020,Nakip2021}, 
to guarantee that in the presence of attacks that create large packet flows, the Server can operate seamlessly and accomplish its role for AD and other useful IoT processing functions,
 \item Experimentally demonstrate that mitigation actions can be triggered to rapidly eliminate the long-term effects of such UDP Flood attacks from the system as a whole.
 \end{itemize}
 
 Thus in Section \ref{Initial}, we provide new measurements on the experimental test-bed shown in Figure \ref{Zero-0},  to illustrate the effect of a UDP Flood attack emanating from an IoT traffic source against the IoT Server that receives packets from different IoT devices. These measurements show that the Server is significantly impacted during an attack and is impeded from conducting its AD functions in a timely fashion.
 
 Based on this observation, Section \ref{QDTP} proposes and evaluates a novel system architecture shown in Figure \ref{Forwarder} where the Server is preceded by a Smart   ``Quasi-Deterministic Forwarding Policy (QDP)'' Forwarder (SQF) that shapes the traffic that is forwarded to the Server. 
 Our results show that if we select the SQF parameters based on mathematical principles \cite{ICC22}, then the SQF effectively limits the undesirable effects of an attack against the Server. However, attack packets accumulate at the SQF which protects the Server, and mitigation actions may discard the accumulated attack traffic.

 \section{Initial Measurement Results} \label{Initial}
 
\begin{figure}[t!]
	\centering
	\includegraphics[height=5cm,width=8cm]{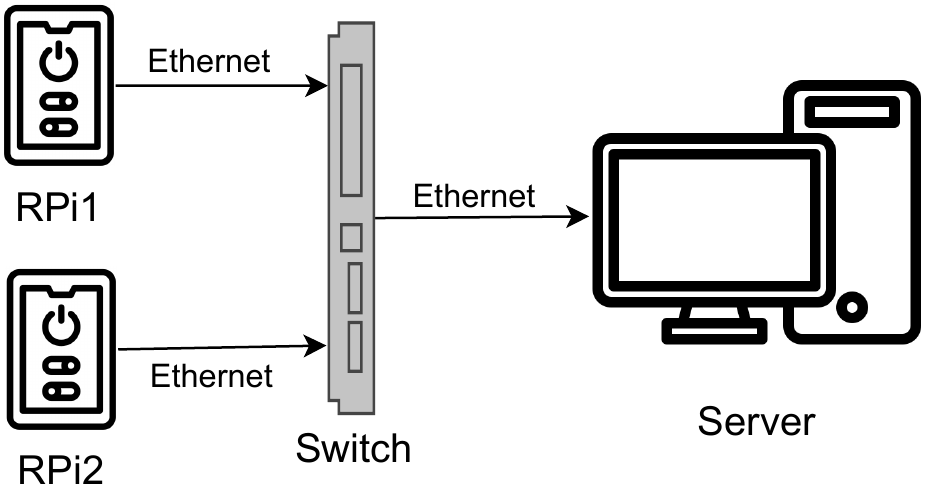}
	\caption{The test-bed composed of an Ethernet network with Raspberry Pi machines that generate
		normal traffic, as well as possible traffic. An IoT Server receives the IoT traffic via the same network.}
	\label{Zero-0}
\end{figure}

As a first step, we have conducted experiments  on the Local Area Network (LAN) test-bed for a system shown in Figure \ref{Zero-0}, in which IoT devices represented by several Raspberry Pi machines, send UDP traffic to the Server. One of the Pi machines is also programmed to generate attack traffic either at predetermined instants or at random. 

These Raspberry Pi $4$ Model B Rev $1.2$ machines (RPi$1$ and RPi$2$) machines, each have a $1.5$GHz ARM Cortex-A$72$ quad-core processor and $2$GB LPDDR$4-3200$ SDRAM, running  Raspbian GNU/Linux $11$ (bullseye), a Debian-based operating system optimized for Raspberry Pi hardware. The normally operating (uncompromised) Raspberry PIs periodically send UDP Protocol packets that simply contain the measurements of the temperature of the Raspberry Pi, to the Server that is shown in Figure \ref{Zero-0}. The choice of this particular data is simply
in order to provide an example of real data that the Raspberry Pis can send.

The Server is an Intel 8-Core i$7-8705$G, with a Linux $5.15.0-60-$ generic $66-$Ubuntu SMP based operating system. It is equipped with $16$GB of RAM, it runs at $3.1$Ghz and has a $500$GB hard drive. It communicates with the Raspberry Pis via the Ethernet Local Area Network (LAN) shown in Figure \ref{Zero-0}, and receives IoT traffic from them via the UDP protocol. 

\begin{figure}[t!]
	\centering
	\includegraphics[height=5cm,width=8.5cm]{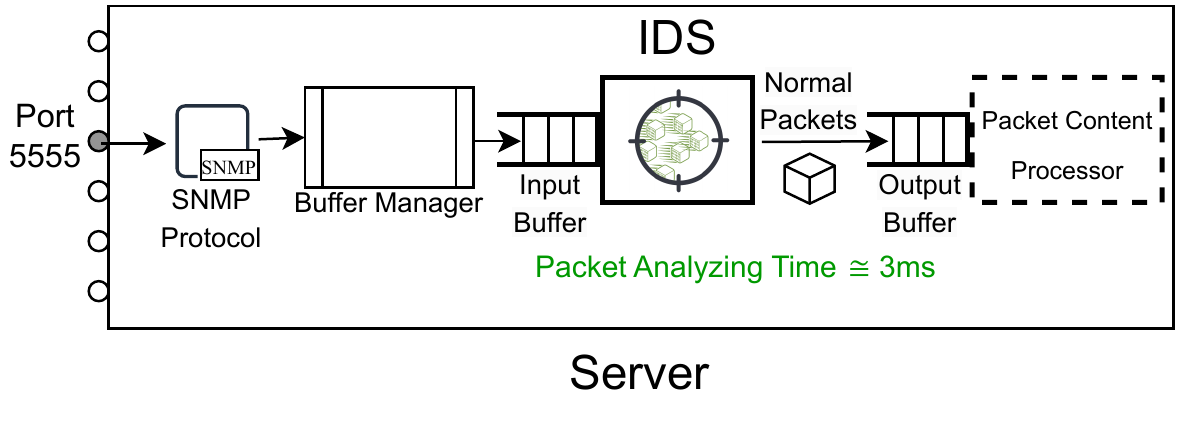}
	\caption{Internal software architecture of the Server, containing a SNMP network protocol manager,
		an AD system \cite{CDIS} that identifies attack packets, and software for processing the  incoming packet contents.}
	\label{Zero-1}
\end{figure}

As shown in Figure \ref{Zero-1}, the Server supports the UDP protocol with SNMP  for incoming packets. It operates the accurate AD algorithm reported in \cite{CDIS}, and
supports the other normal processing needs of incoming UDP packets.  The  UDP protocol's simplicity fits the needs of the simple IoT devices that we use, since UDP does not establish a connection before transmitting and does not use ACKs or error recovery for communications \cite{kumar2012survey}. 

While many datasets can be used to generate attack traffic, including
the KDD$99$ dataset or its improved version NSL-KDD,  UNSW-NB15, CICDS$2017$, and the Bot-IoT dataset \cite{arthi2021design}, in this work we use MHDDoS \cite{MHDDOS} containing $56$ recent real-world DoS attacks with $56$ different techniques, for attack traffic emulation in this Ethernet based test-bed.

\begin{figure}[t!]
 	\centering
 	\includegraphics[height=5.7cm,width=9.5cm]{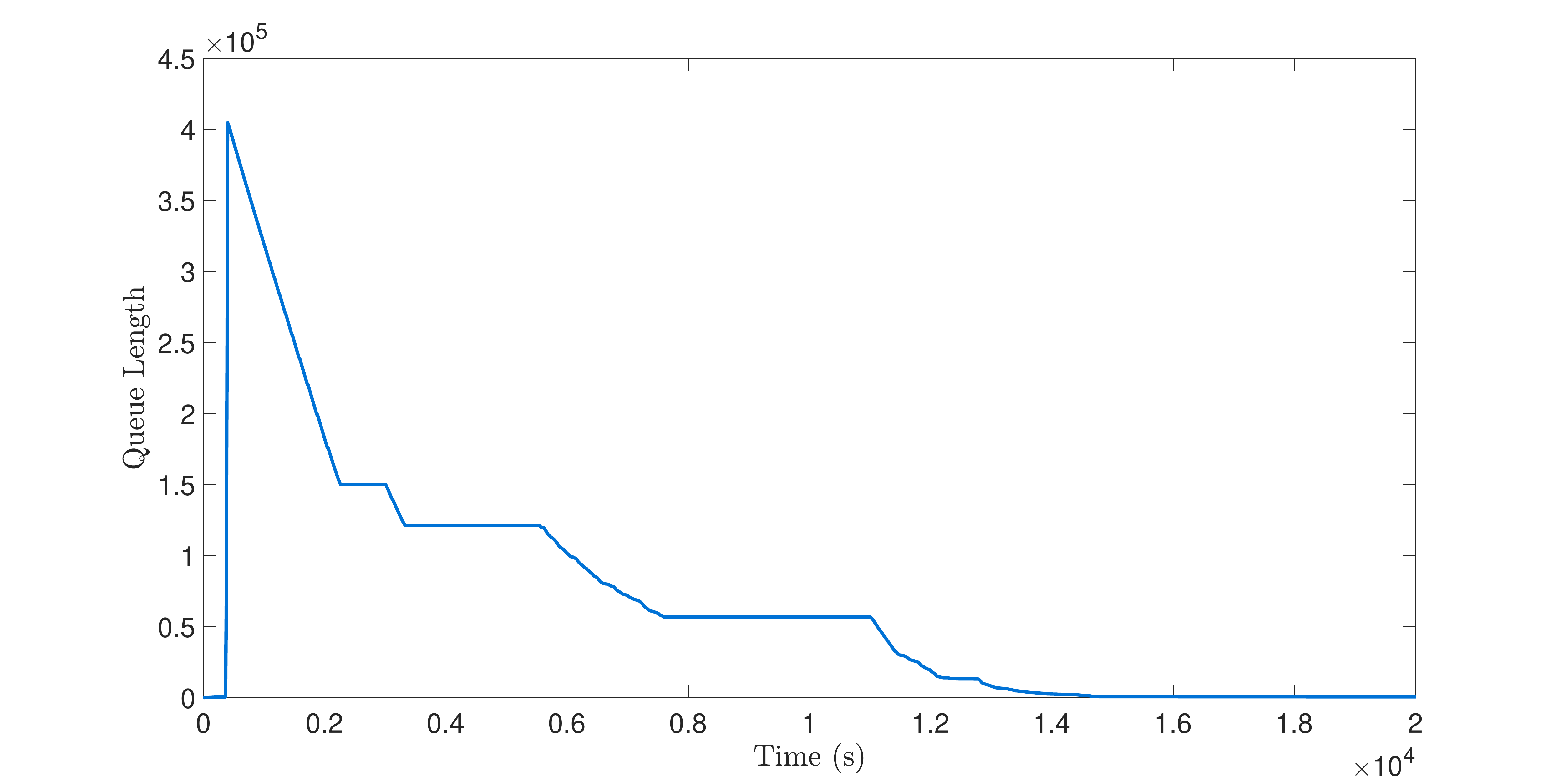}
 	\caption{Experimentally measured queue length (the $y$-axis is in number of packets) over time (the $x$-axis is in seconds) at the Server input prior to processing at the AD module,
 		during a $60$ second UDP Flood Attack launched from one of the Rasberry Pis of Figure \ref{Zero-0} against the Server. The backlog of packets at the Server
 	initially rises rapidly to $400,000$ packets, and without human intervention
 	the congestion at the Server lasts far longer than the attack itself, up to several hours, due to the fact that the Server is paralyzed and stops its AD processing packets for long time intervals. These long interruptions in AD processing time are observed as the large outliers in AD processing times in Figure \ref{fig:ProcessingTime2}.}
 	\label{Zero}
 \end{figure}

In Figure \ref{Zero} we show measurements of the effect of a $60$ sec Flood Attack, which overwhelms the 8-Core Server with some $400,000$ packets that accumulate at its input buffer. The Server's activities, including AD, are paralyzed by the attack and the packet backlog takes nearly $300$ minutes to clear. Thus we see that the attack significantly impairs its capability for AD, its ability to discard attack packets and to subsequently process benign packets.

\begin{figure}[h!]
	\centering
	\includegraphics[height=5.7cm,width=8.5cm]{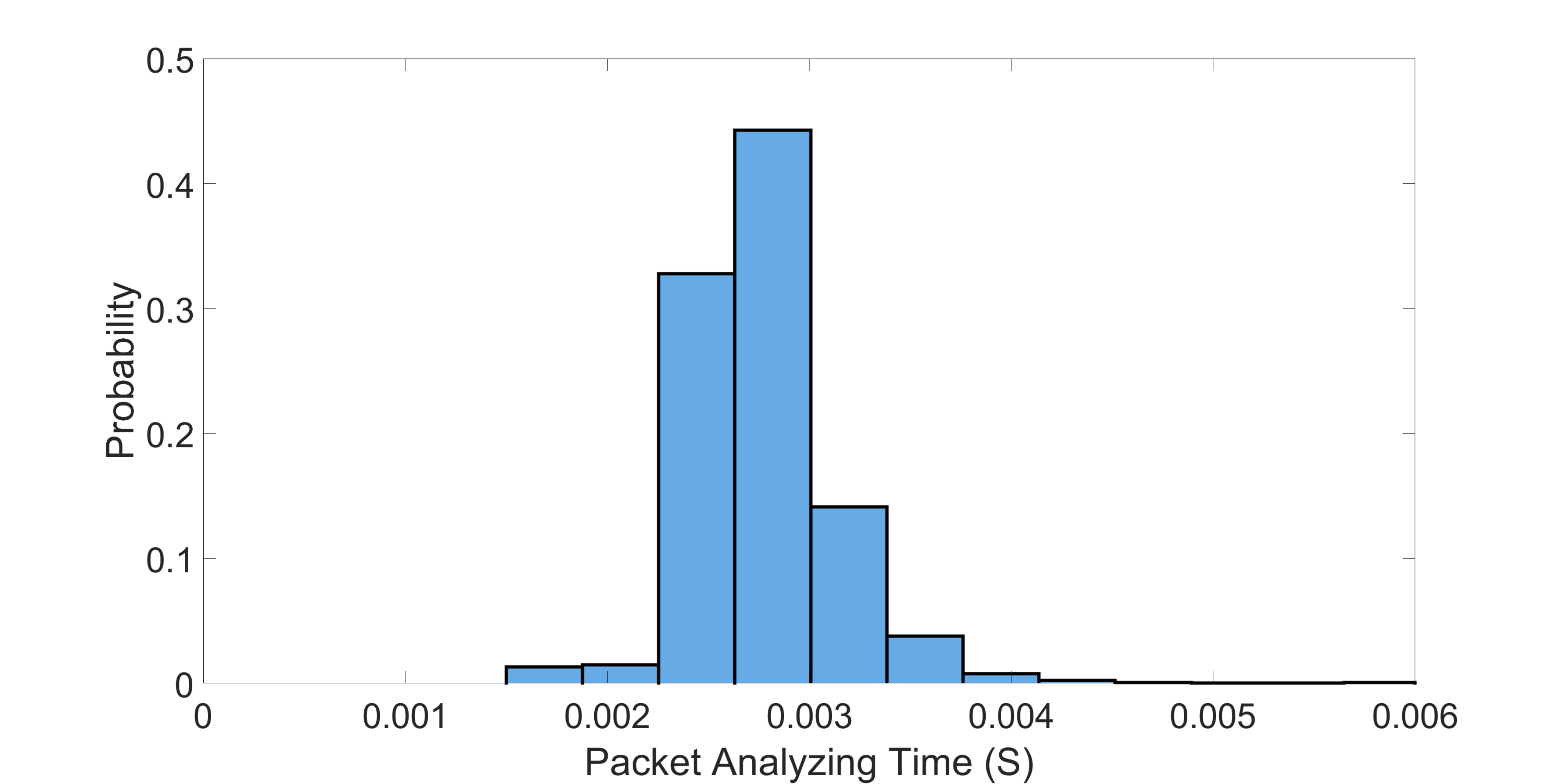}\\
	\centering
	\includegraphics[height=5.7cm,width=8.5cm]{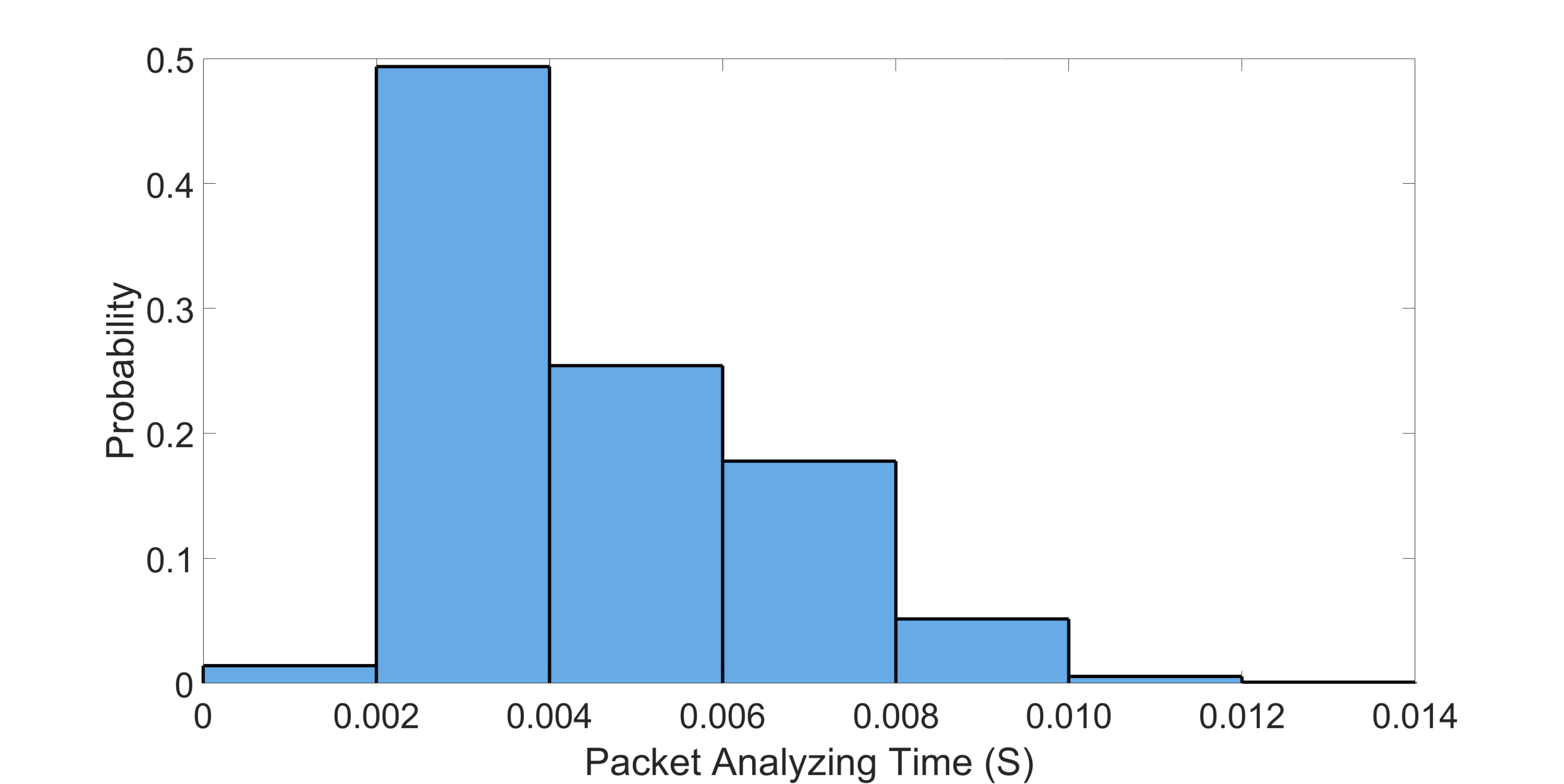}
	\caption{In the Upper figure, we show the histogram of measurements of the Server's AD processing time per packet, when there is no attack, exhibiting an average processing time of $2.98$ ms and variance $0.0055~ms^2$. In the Lower figure an attack is occurring: the Server's 
		measured average AD processing time of packets rises substantially 
		to $4.82$ ms with a variance  $0.51~ms^2$, }
\label{fig:PT1}	
\end{figure}

\begin{figure}[h!]
	\centering
	\includegraphics[height=5.7cm,width=9cm]{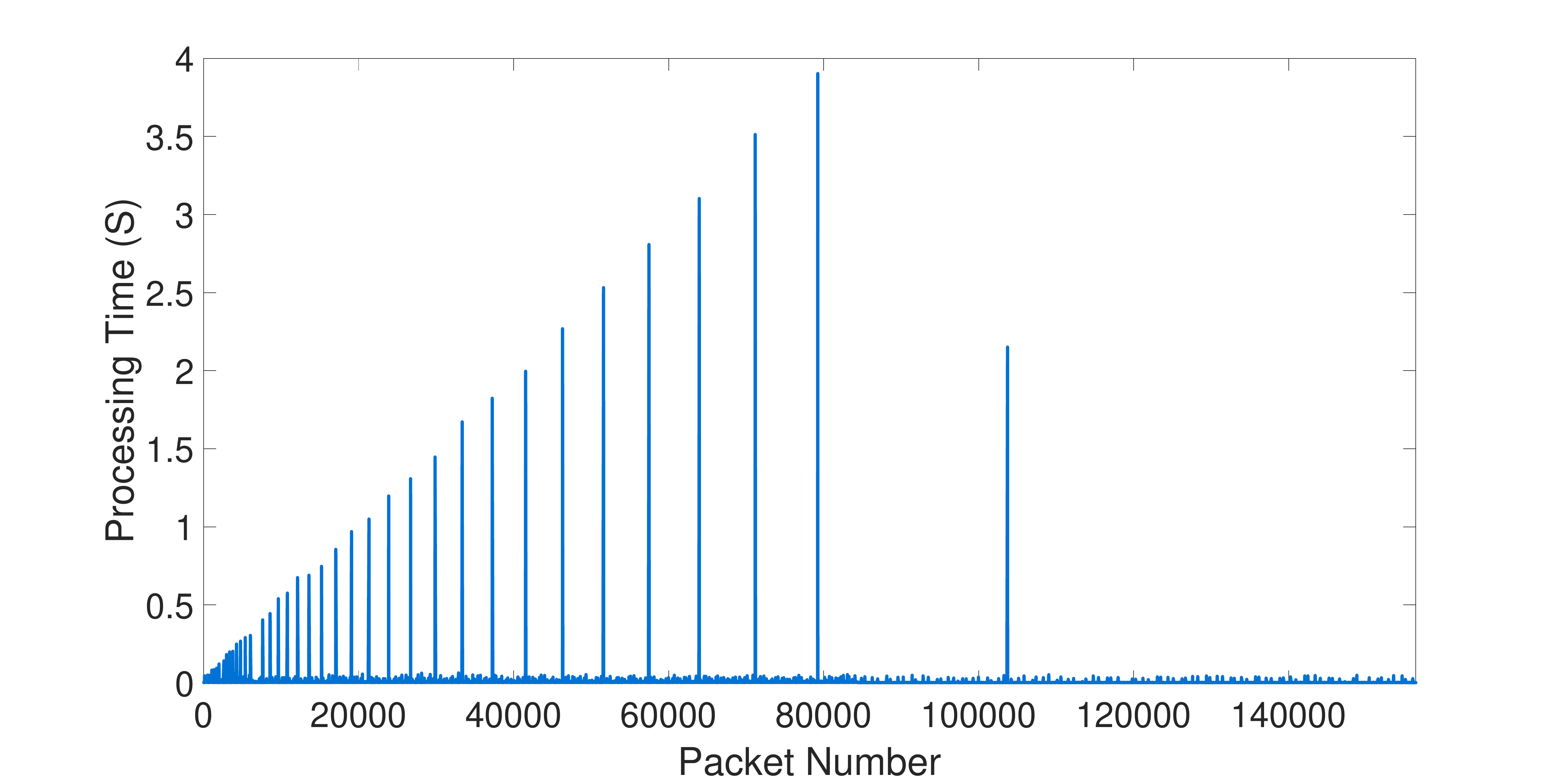}\\
	\centering
	\includegraphics[height=5.7cm,width=9cm]{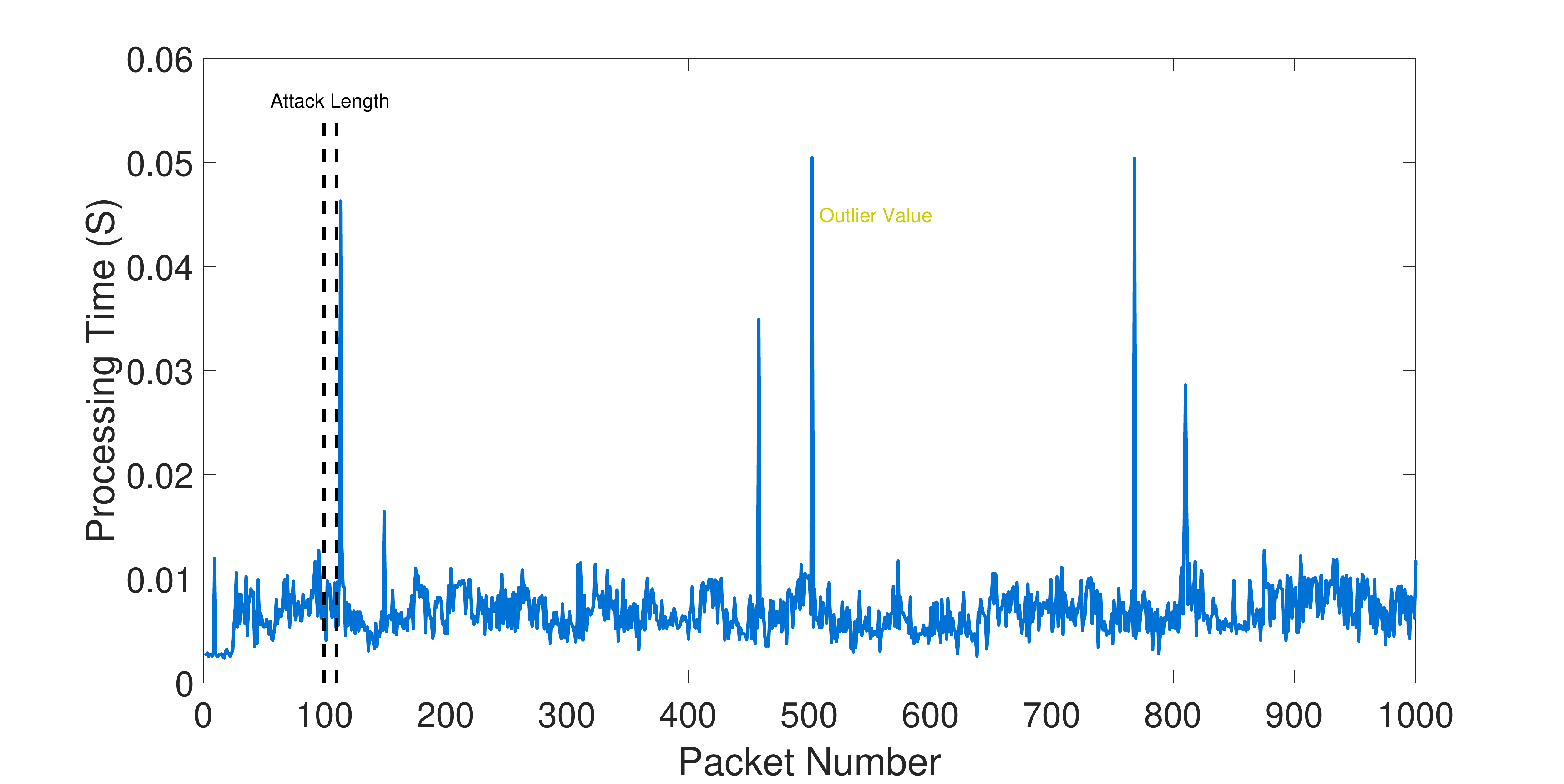}
	\caption{In the Upper figure, we show successive measurements of the Server's AD processing time per packet during a UDP Flood Attack (in the absence of the QDTP Forwarder SQF), showing large outliers that initially become more severe,  and gradually become less frequent over time. In the Lower figure, the AD processing time of packets that is measured after the UDP Attack begins, reveals very large outliers in AD processing times,  indicating that the AD is intermittently paralyzed or unable to operate. }
	\label{fig:ProcessingTime2}
\end{figure}

\begin{figure}[t!]
 	\centering
 	\includegraphics[height=5.5cm,width=8.5cm]{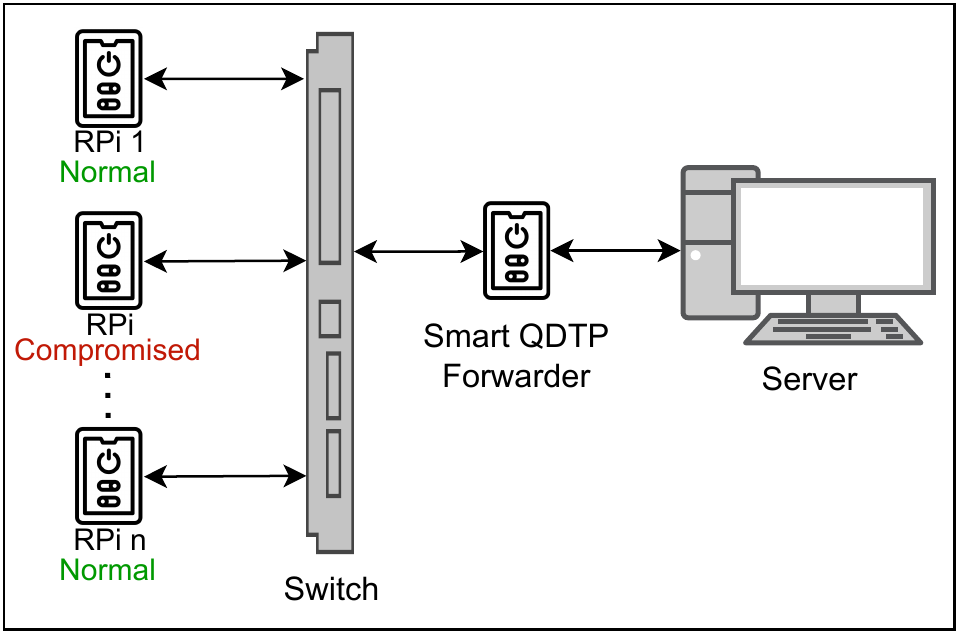}
 	\caption{The figure shows the modified system architecture where a Smart QDTP Forwarder (SQF) constantly acts as a traffic
 	shaping interface between the Ethernet LAN and the Server. The effect of the
 	SQF is to eliminate the paralyzing effect of the
 	packet flood at the Server, buffering packets within the SQF and forwarding in a manner which allows the Server to conduct its AD processing and other work in a timely fashion.}
 	\label{Forwarder}
 \end{figure}

The detailed measurements of the Server's AD processing times per packet, when there is no attack, and when a UDP Flood Attack occurs,  are reported  in Figure \ref{fig:PT1}
and Figure \ref{fig:ProcessingTime2}. 

We observe that the Server's AD processing time per packet, when {\bf no attacks} occur,  has an average value of $2.98$ milliseconds (ms). 
On the other hand, when the Server is targeted by a UDP Flood Attack,
we observe a substantial increase in the AD algorithm's average processing time to $4.82$ ms. 
Moreover, the AD processing time per packet when the Server is under attack, exhibits some very large outliers, as shown in Figure \ref{fig:ProcessingTime2}. We observe 
that these ``outlier'' processing times are close to $10^3$ times larger than the typical values, showing that during a UDP Flood Attack the Server's AD procesing of packets  is repeatedly paralyzed
for a substantial amount of time, as also shown in Figure \ref{Zero}. 

\begin{figure}[h!]
	\centering
	\includegraphics[height=5.7cm,width=8.5cm]{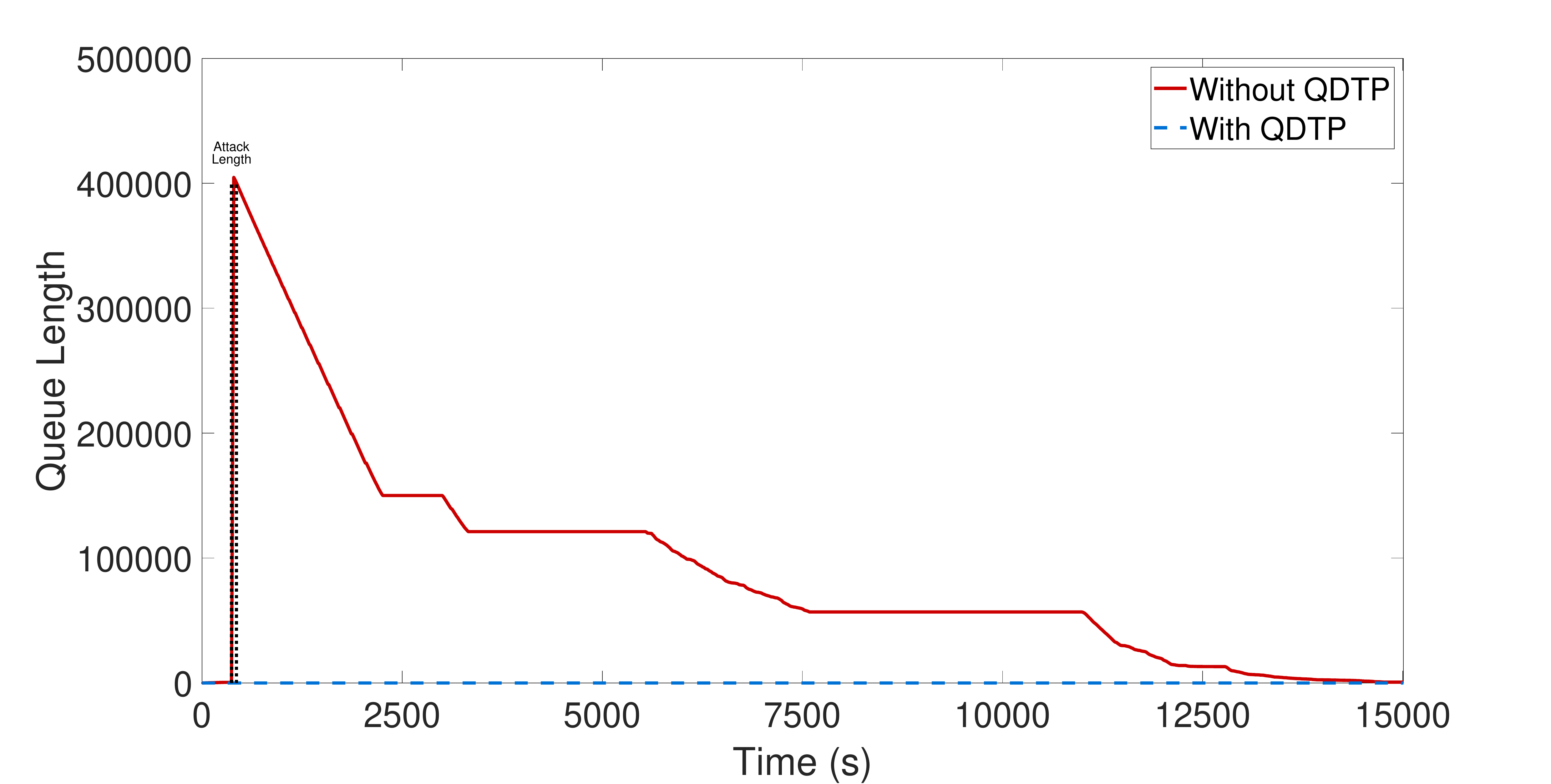}\\
	\centering
	\includegraphics[height=5.7cm,width=8.5cm]{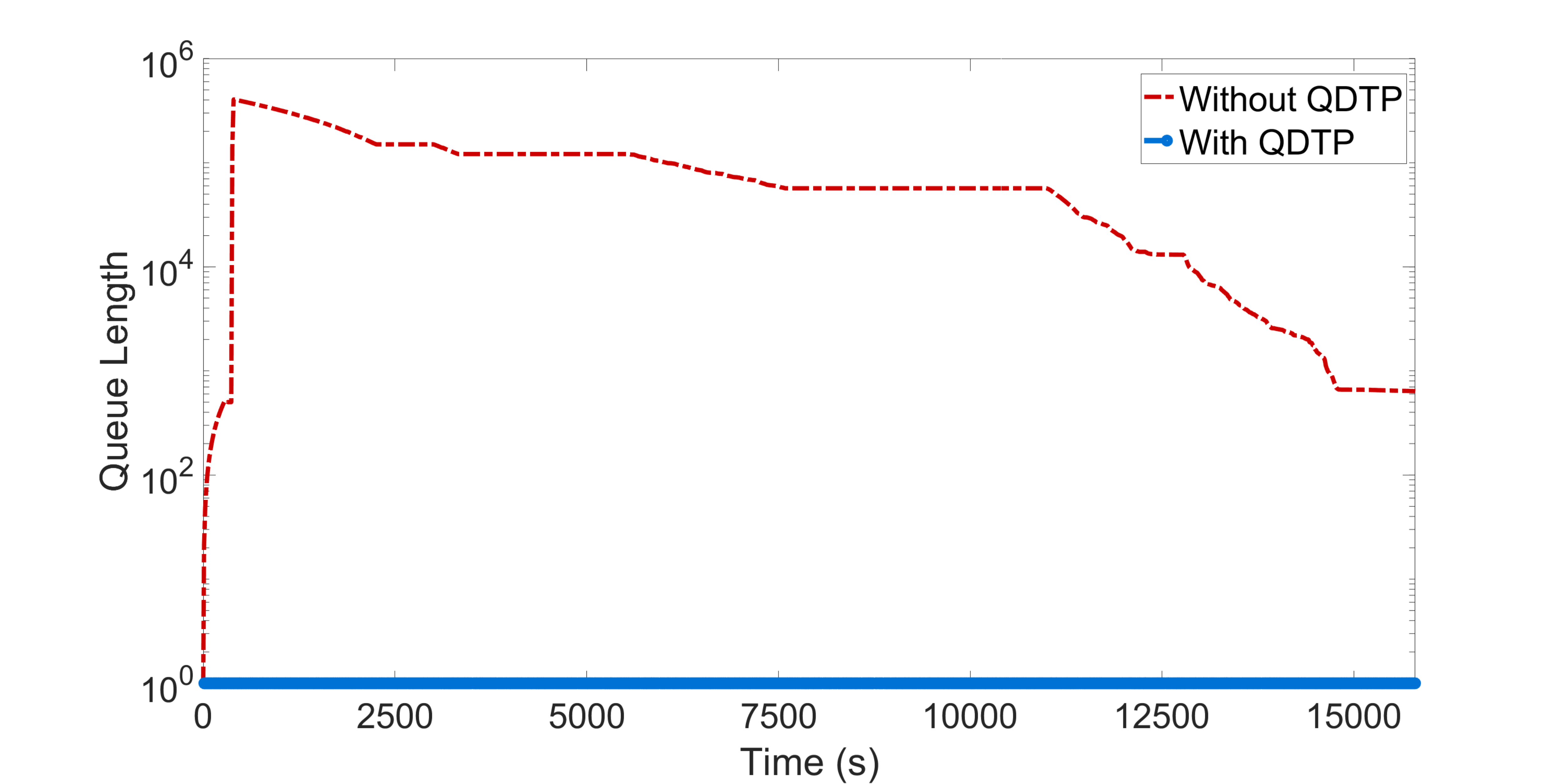}
	\caption{The queue length at the Server in the presence of a $60$ second UDP Flood Attack. The  figure Above shows the case {\bf without}  the SQF, and we see that the queue length peaks to $400,000$ packets and descends slowly over some $15,000$ seconds. The figure Below compares the 
		queue length in logarithmic scale, {\bf with SQF
			in Blue} using the parameter $D=3~ms$, against the case {\bf without SQF in Red}, with the same UDP Flood Attack which lasts $60$ seconds. Note that because  the value of $D$ we use is very close to the
		average value of $T_n$ measured to be $2.98$ ms in the absence of an attack, as shown in Figure \label{fig:PT}, the fluctuations in the values of $T_n$ will cause a small queue buildup (in the order of a few packets), as seen  in te Blue plot in the figure Below.} 
	\label{QL-60s}
\end{figure}

\begin{figure}[h!]
	\centering
	\includegraphics[height=5.7cm,width=9cm]{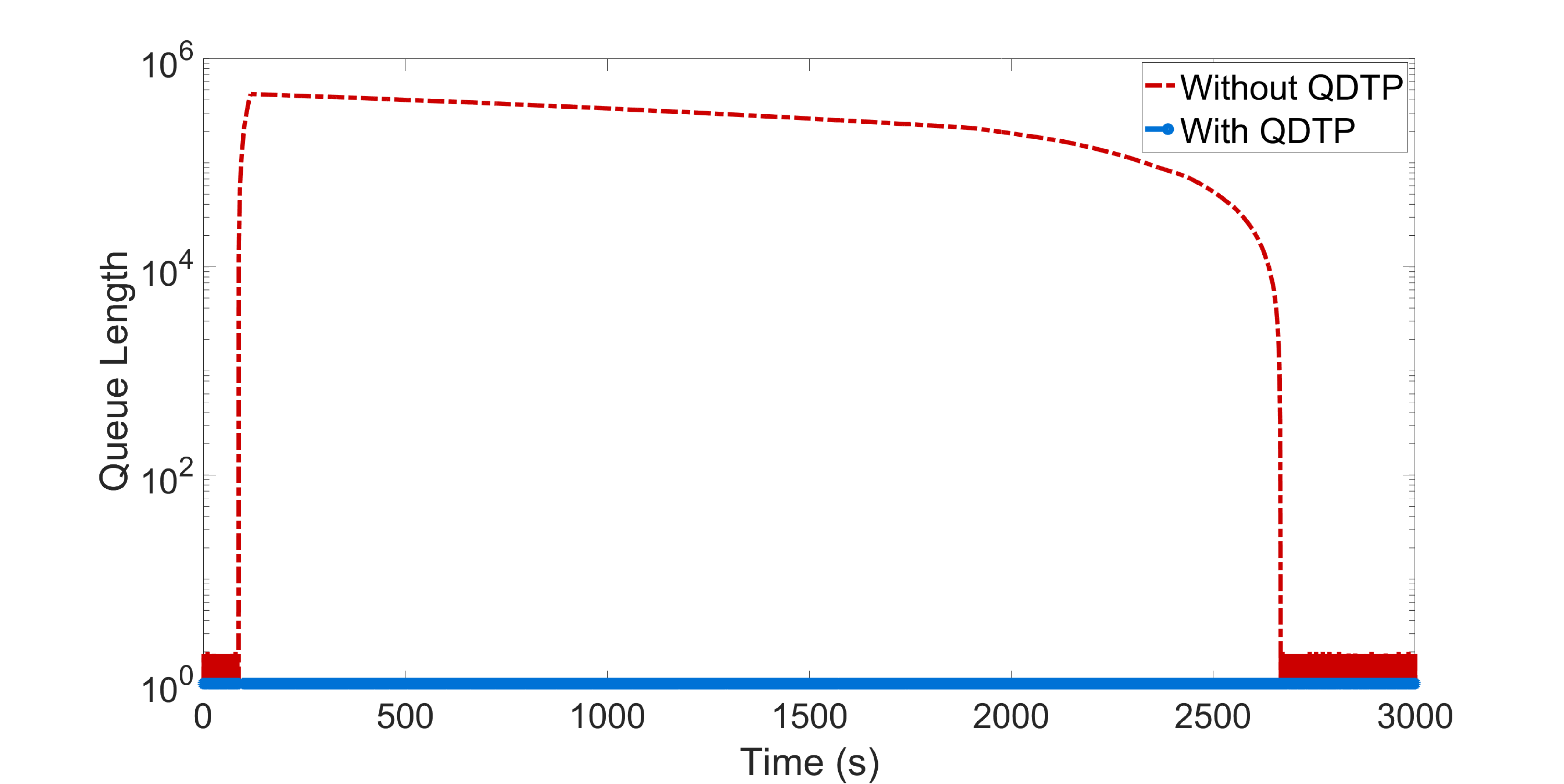}\\
	\includegraphics[height=5.7cm,width=9cm]{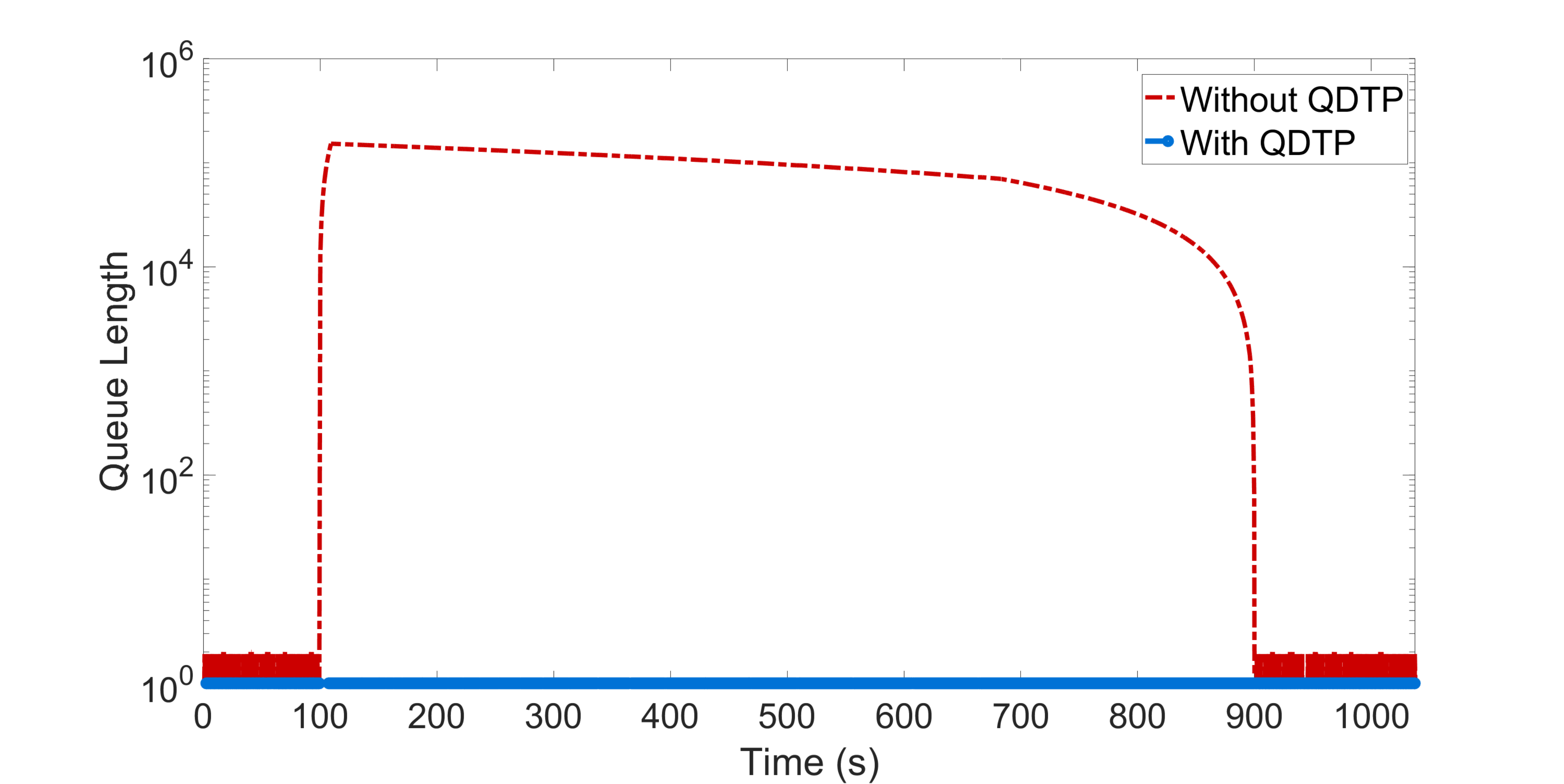}
	\caption{We measure the Server queue length (repesented logarithmically)  when the Server is targeted by a UDP Flood Attack that lasts $30-sec$ (Above) and $10-sec$ (Below). The Red curves show  the case without the SQF traffic shaping Forwarder, while the Blue Curves
		show the effect of the use of the SQF which uses QDTP. We observe the huge difference in queue length.
		For both the $30$ and $10$ second attacks, we have set $D =3 ~ms$.}
	\label{QL-30s}
\end{figure}

\begin{figure}[t!]
	\centering
	\includegraphics[height=4cm,width=8.5cm]{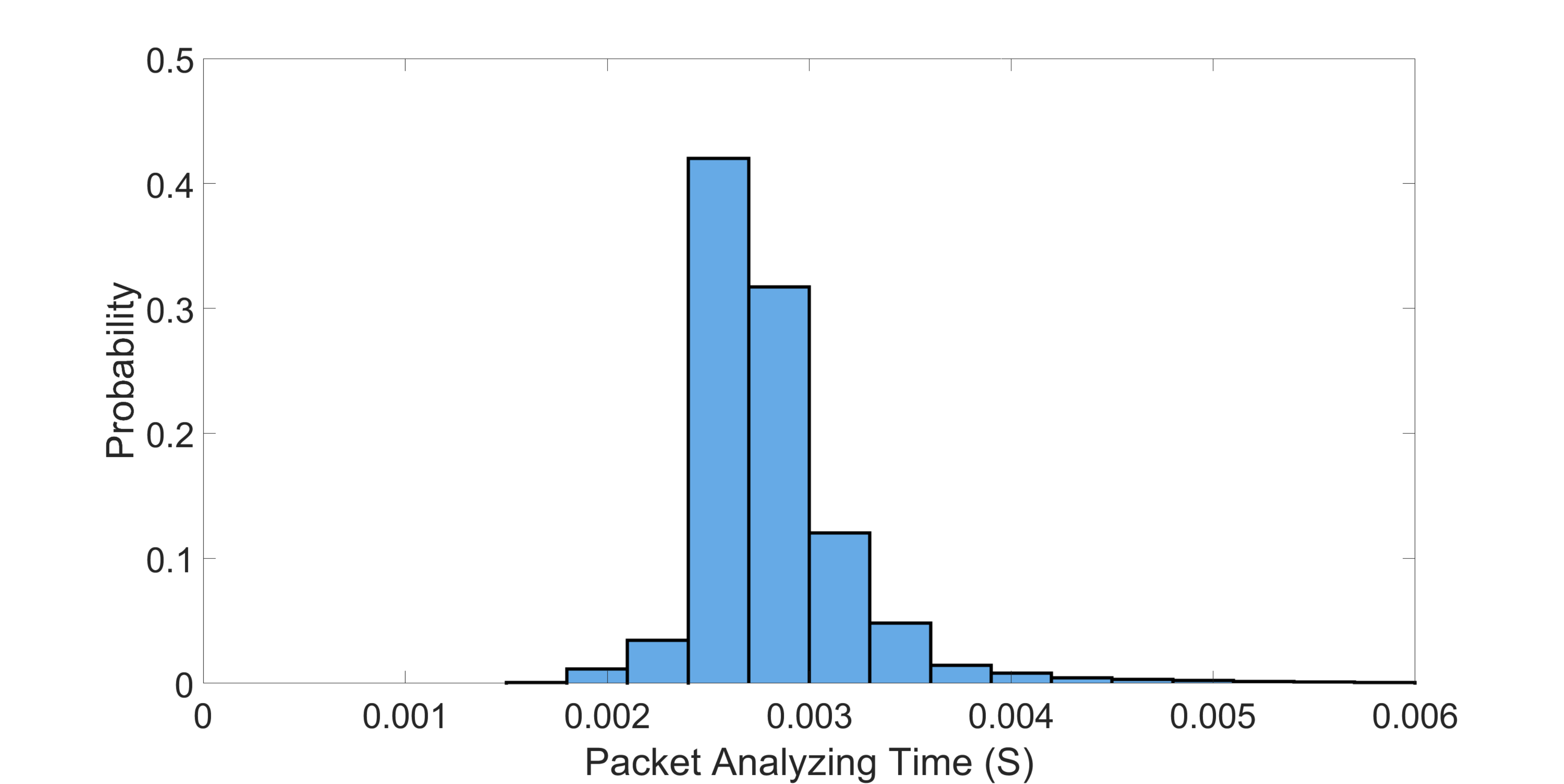}\\
	\includegraphics[height=4cm,width=8.5cm]{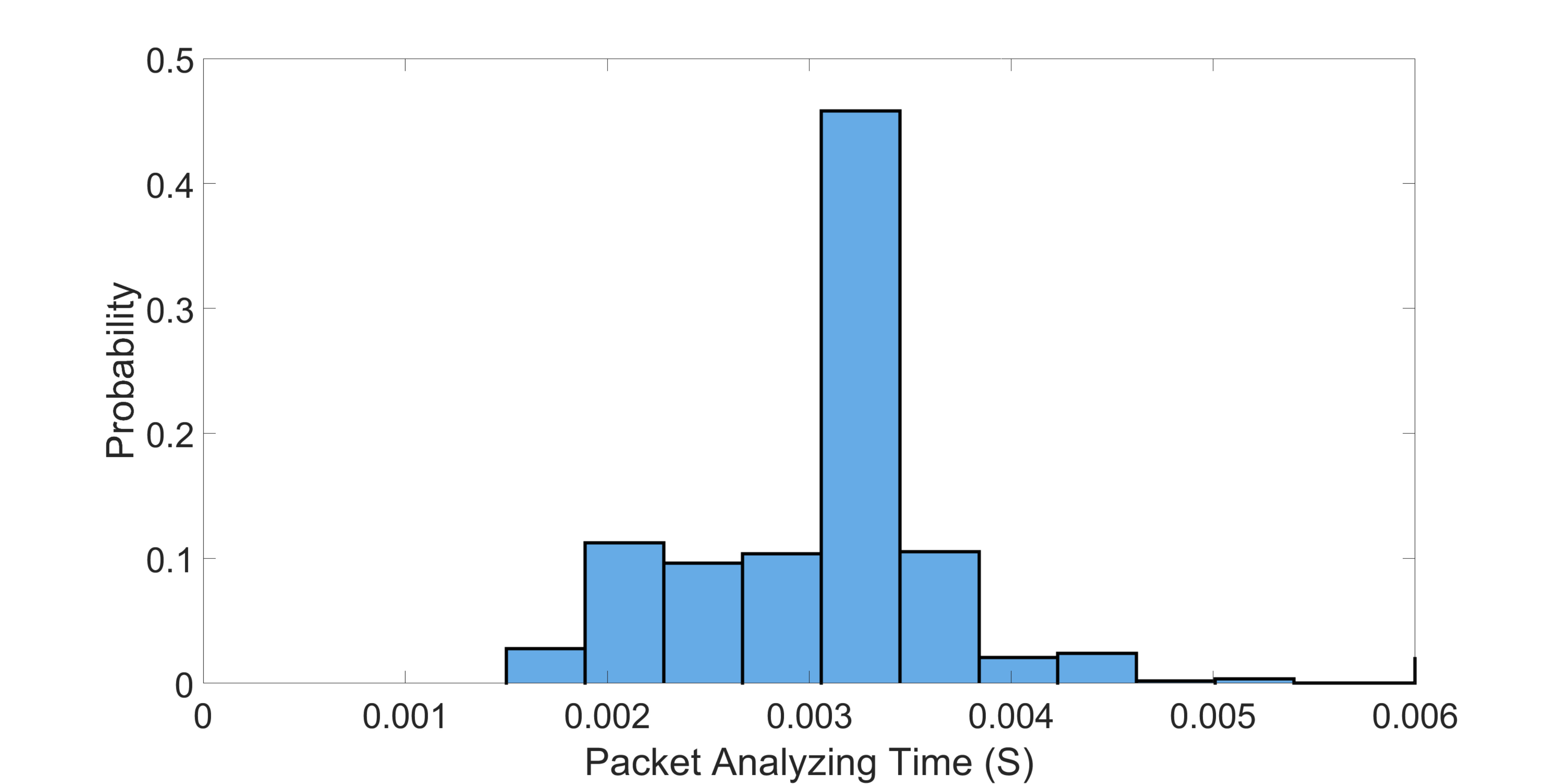}
	\caption{AD Processing Time at the Server when the SQF with the QDTP Policy is installed and the parameter $D=2.7~ms$ is used. We observe that the AD processing time $T_n$ has an average value of
		$2.97~ms$ and variance of 	$0.0041~sec^2$ in the absence of an attack (Above). In the presence of a UDP Flood Attack (Below) the average processing time of the AD per packet is higher by roughly 10\% on average, at $3.28~ms$ with a variance of  $0.0023~sec^2$ so that the SQF is effective in protecting the Server from paralysis and excessive slowdown.}
	\label{Fig4}
\end{figure}

\begin{figure}[t!]
	\centering
	\includegraphics[height=4cm,width=8.5cm]{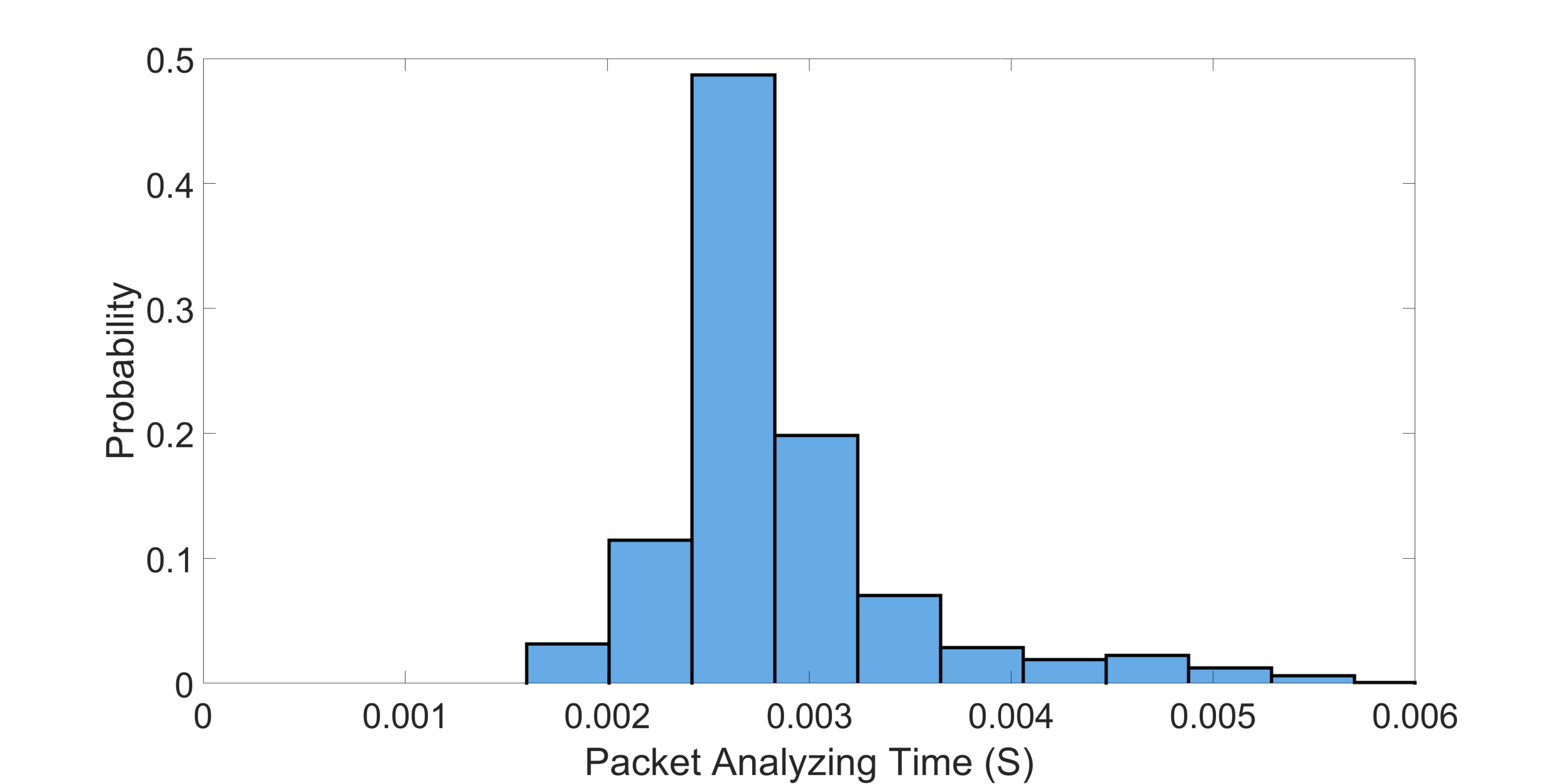}\\
	\includegraphics[height=4cm,width=8.5cm]{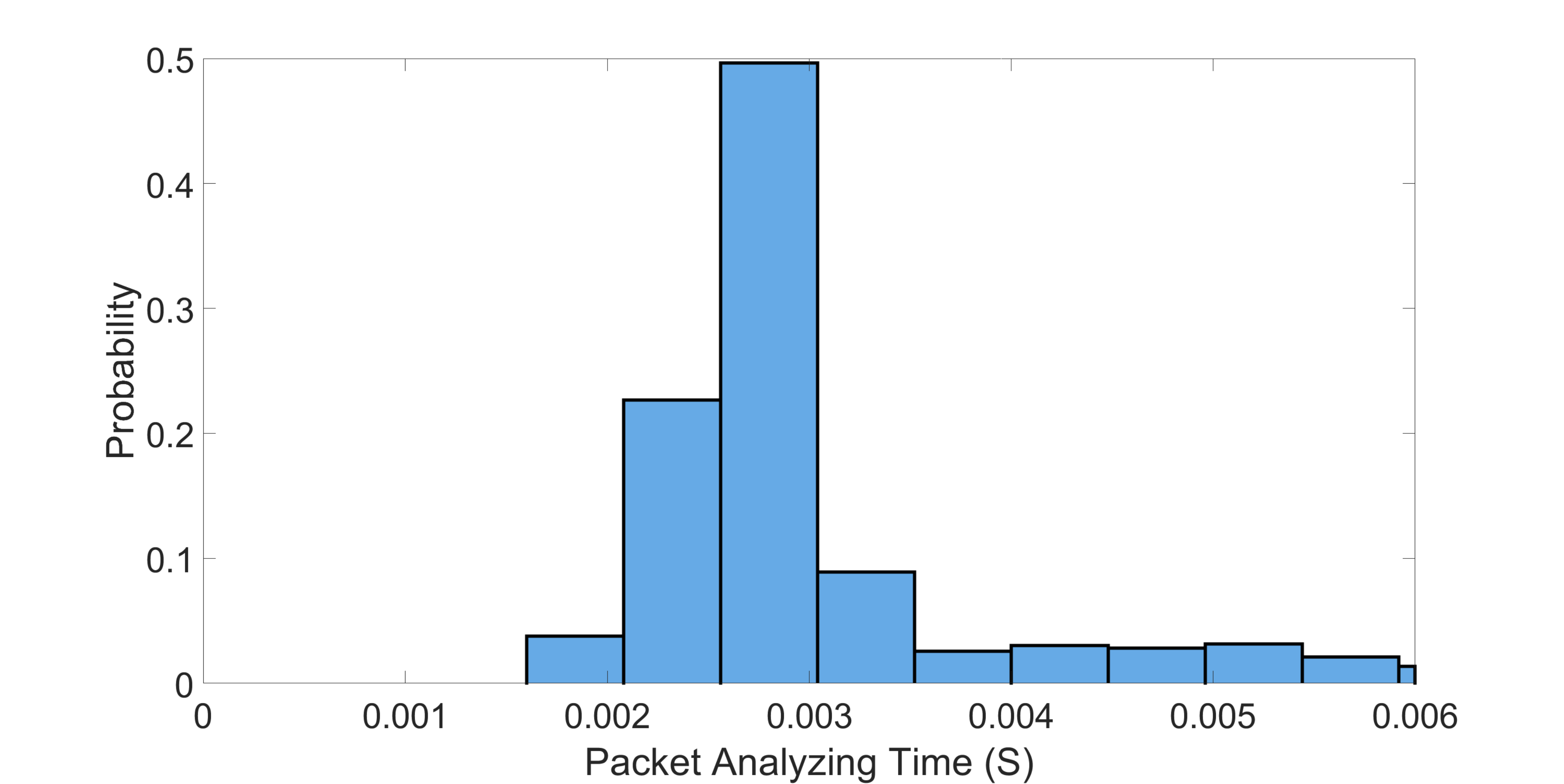}
	\caption{AD Processing Time at the Server when the SQF with the QDTP Policy is installed and the parameter $D=3.20~ms$ is used. We observe that the AD processing time $T_n$ has an average value of
		$3.00~ms$ and variance of 	$0.0036~sec^2$ in the absence of an attack (Above). In the presence of a UDP Flood Attack (Below) the average processing time of the AD per packet is quasi-identical on average, at $2.99~ms$ with a variance of  $0.0067~sec^2$ so that in this case too, the SQF is effective in protecting the Server from paralysis and excessive slowdowns.}
	\label{Fig5}
\end{figure}

\begin{figure}[h!]
	\centering
	\includegraphics[height=5.7cm,width=9cm]{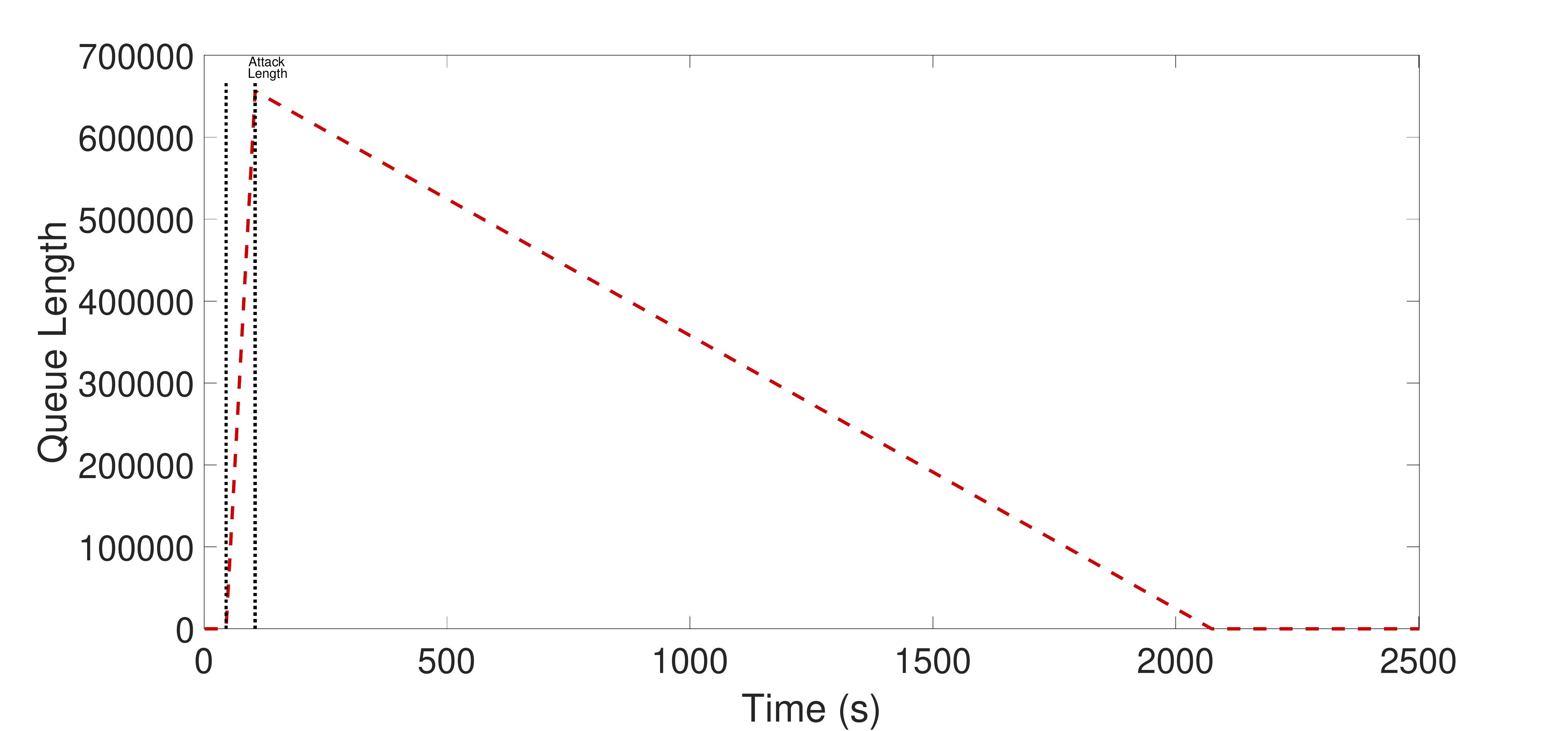}
\caption{SQF queue length ($y$-axis in number of packets) against time ($x$-axis in seconds) when a UDP Flood attack lasts for $60$ seconds. We have used $D =3~ms$, and no mitigation action takes place.}
	\label{fig:RPI-QL-30s-60s}
\end{figure}

\subsection{Lindley's Equation when the SQF is not Used}

 If  the SQF module is {\bf not} being used as shown in Figure \ref{Zero-0}:
 \begin{itemize}
 	\item  Let $0=a_0\le a_1\le a_2,~ ...~$, be the successive packet arrival instants at the Server  through the Ethernet LAN from any of the IoT devices connected to the LAN. We also define the interarrival time $A_{n+1}=a_{n+1}-a_n$. 
 	\item Let $T_n$  denote the Server's AD processing time for the $n-th$ packet, and assume that
 	the Server processes packets in First Come First Served (FCFS) order.
\end{itemize} 
Then the total waiting time $L_{n+1}$ of the $n+1$-th incoming packet, between the instant $a_n$ and the start of the AD processing time of the Server, is given by the well known Lindley's equation \cite{Sigman,Gelenbe}:
 	\begin{equation}
 	L_{n+1}=[L_n+T_n-A_{n+1}]^+,~n\geq 0,~L_0=0~,\label{Lindley}
 	\end{equation}
 	where for a real number $X$, we use the  notation:
 	\begin{equation}
 	[X]^+=X~if~X>0,~and~ [X]^+=0~if~X\leq 0.
 	\end{equation}
 	Note that $L_0=0$ because the first incoming packet encounters an empty queue infront of the AD.
 	Note also that whenever we have $T_n>A_{n+1}$ then $L_{n+1}>L_n$, i.e. the waiting time increases.
 	
During a Flood Attack, the values of $A_n$ and $T_n$ will be modified, as we see from
Figure \ref{Zero}, indicating that packet arrival rates have considerably increased so that the values of $A_n$ are much smaller,
while Figure  \ref{fig:PT1} shows that the values of $T_n$ are also larger. However the form of (\ref{Lindley}) does not change.

  \section{Effect of the Smart QDTP Forwarder (SQF) } \label{QDTP}
  
 In Figure 	\ref{Forwarder}, we present our proposed modified architecture where the Server, whose role is to 
 process incoming IoT packets -- including operating the AD module in order to detect attacks -- is ``protected'' by a Smart QDTP Forwarder (SQF) which is placed between the Ethernet based sources of IoT traffic, and the Server's input port.   The role of the  SDF is to shape the incoming traffic directed at the Server using the {\em Quasi-Deterministic Transmission Policy (QDTP)}  \cite{Mascots21,ICC22}. 
 
 QDTP is a simple policy that delays some of the packets it receives, by forwarding them to the Server at time $t_n \geq a_n$, where $a_n$ is the $n$-th packet's arrival instant to the SQF, and $t_n$ is the instant at which SQF forwards the packet to the Server, and is defined by:
\begin{eqnarray}
&&t_{n+1}=\max\{t_n+D, a_{n+1}\},~t_0=a_0,~n\geq 0,\label{eq1}\\
&&so~that~t_{n+1}-t_n\geq D~,\label{eq2}
\end{eqnarray}
where $D>0$ is a constant parameter of the QDTP algorithm that needs to be fixed.

When the $n-th$ packet is transmitted by the SQF, we assume that it arrives instantaneously at the Server's input queue for AD processing. Here we are in fact assuming that the physical transmission time from the SQF to the Server, and the network protocol service time inside the Server, are tiny compared to the AD processing duration $T_n$ at the Server. Thus the total delay $Q_n$ experienced by the $n$-th packet due to the action of the SQF, that elapses from the arrival of the $n$-th packet to the SQF at $a_n$, until its arrival to
  	the AD at the Server at $t_n$, is:
  \begin{eqnarray}
&&Q_0=t_0-a_0=0,\label{Q0}\\
&&Q_{n+1}=t_{n+1}-a_{n+1},\nonumber\\
&&=\max\{t_n+D,a_{n+1}\}-a_{n+1},\nonumber\\
&&=0,~if~t_n+D\leq a_{n+1},~and\nonumber\\
&&=t_n+D-a_{n+1},~otherwise.
  \end{eqnarray}
Since $t_n=Q_n+a_n$, we obtain the recursive expression:
\begin{eqnarray}
&&Q_{n+1}=[t_n+D-a_{n+1}]^+,\nonumber\\
&&=[Q_n+D-A_{n+1}]^+,~n\geq 0,\label{Q}
\end{eqnarray}
which is also an instance of Lindley's equation (\ref{Lindley}).

On the other hand, the Server's AD module also acts as a FCFS queue and we can exploit
Lindley's equation again to compute $W_n,~n\geq 0$ the waiting time of the $n$-th packet that arrives to the Server to be processed for attack detection, which is:
\begin{eqnarray}
W_{n+1}&=&[W_n+T_n-(t_{n+1}-t_n)]^+,~W_0=0,\label{Server1}\\
&\leq &W_n+T_n-(t_n-t_{n+1}), \label{Server2}
\end{eqnarray}
since the $n$-th packet's AD service time is $T_n$ and the $n+1$-th interarrival interval to the
Server's AD queue is $t_{n+1}-t_n$.

Therefore using  equations (\ref{Server2}) and (\ref{eq2}) we obtain:
\begin{equation} 
W_{n+1}\leq W_n + T_n - D, \label{Server3}
\end{equation}
and we have the following key insight into how to choose $D$:

\medskip
\noindent{\bf Result 1.} If we fix the parameter $D$ in the QDTP policy for the SQF to a value so
that $D>T_n$ for all $n\geq 0$, then the waiting time $W_n$ at the Server will remain at the value $W_n=0$ for all $n\geq 0$.

\medskip

We now present experiments showing the usefulness of Result 1.
Noting from Figure \ref{fig:PT1} that the measured average value of $T_n$ is $2.98~ms$ when there is no attack,
we first select $D=3~ms$ which is just above that value.

Figure \ref{QL-60s} compares the case {\bf without SQF} (Above) and {\bf with SQF} (Below) during a $60~sec$ UDP Flood Attack. Note that the figure Above represents the Server queue length varying over time, without the SQF. The figure Below is in logarithmic scale for the Server queue length, and compares the cases without SQF (in Red) and with SQF (in Blue) for the  Server queue length varying over time. Since $D=3~ms$ is very close to the
average of $T_n$, the fluctuations in the values of $T_n$ cause a small queue buildup 
of a few packets, as seen  in the Blue plot in the lower part of the shown Below. 

Figure 		\ref{QL-30s} shows the results of four experiments where we measure the queue length at the Server when a UDP Flood Attack lasts $30$ (Above) and $10$ (Below) seconds, without (Red) and with (Blue) the SQF Forwarder. Without SQF, the Server's AD processing time  increases significantly.
In the $30~sec$ attack, approximately $470,000$ packets are received at the Server and without SQF it takes $44.45$ minutes for the Server to return to normal process them, while in the $10~sec$ attack $153,667$ packets are received
and it takes the Server roughly $15$ minutes to process them. Note that in these curves it takes
some $99$ seconds for the compromised RPi to launch the attack.

Figure \ref{Fig4} shows that  when we use the SQF based system  with $D=2.7~ms$, which is 
smaller than the value recommended by  Result 1,  when there is no attack this choice of $D$ 
has very little effect. However, when a UDP Flood Attack occurs, the Server's AD processing is somewhat slowed down and the average value of $T_n$ increases by roughly 10\%.

On the other hand, Figure \ref{Fig5} confirms {\bf Result 1}  since it shows that, if we take $D=3.2~ms$ which guarantees that $D>T_n$ most of the time, then the measured average value of $T_n$ remains at around $3~ms$ showing that it has not been slowed down by the attack's overload effect. Of course the same is seen when no attack occurs.

\section{SQF Queue Buildup and Attack Mitigation} \label{Mitigate}

When a Flood Attack occurs,  the SQF accumulates packets in its input queue,  and forwards them to the Server using the QDTP algorithm with $D=3~ms$, so that the Server does not experience any AD slowdown, ensuring that the Server continues to operate as usual. Figure \ref{fig:RPI-QL-30s-60s} shows the sudden increase and then slow decrease of the SQF input queue  when a UDP Flood attack lasts for $60$ seconds, and 
the SQF uses $D =3~ms$. Since both the SQF (and the Server) do not drop packets, the attack packets will accumulate at the input queue of the SQF.

Thus in this section we test a possible mitigating action that the SQF can take. 
Since Flood Attacks are characterized by an unusually high packet arrival rate, and this is also one of the attack detection parameters used by the AD used in this work \cite{CDIS}, we now test an additional feature,
as follows:
\begin{enumerate}
	\item If the SQF receives more than $N$ packets in a time interval that is less than or equal to $D$,
	it drops all incoming packets for the next $K.D$ time units.
	\item Here $N$ and $K$ are parameters of the mitigating action.
	\item The action is repeated as long as the condition 1) (above) on $N$ persists.
\end{enumerate}
To illustrate the effect of this simple policy, in a first experiment we set $N=10$ and $K=3$ and implement the suggested drop-based mitigation policy.

\begin{figure}[h!]
	\centering
	\includegraphics[height=5.7cm,width=8.5cm]{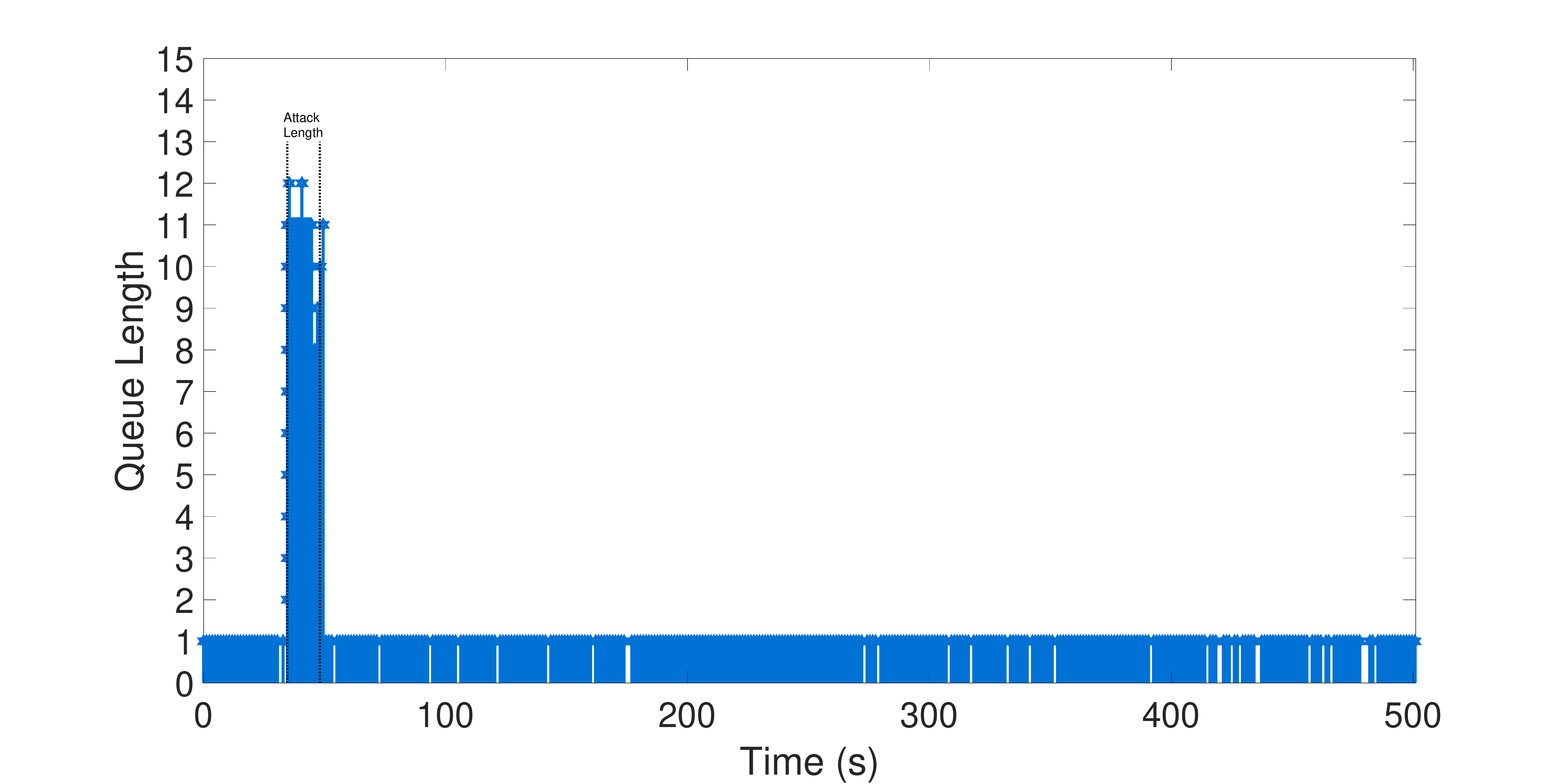}
	\caption{SQF Smart QDTP Forwarder queue length when the attack lasts for $10$ seconds, with  the mitigation action, and $D = ~3 ms$.}
	\label{fig:QueueLengthMitigation10}
\end{figure}

In this experiment, an RPi  launches a $10$ second Flood Attack, and the resulting queue length at the input of the SQF is shown in Figure \ref{fig:QueueLengthMitigation10}, where we see that the SQF input buffer reaches a small value of $12$ packets. The attack starts at the $34$-th second and lasts $10$ seconds, but thanks to the mitigation policy there is no accumulation of packets. After the attack ends the SQF can continue to operate normally.

\begin{figure}[h!]
	\centering
	\includegraphics[height=5.7cm,width=8.5cm]{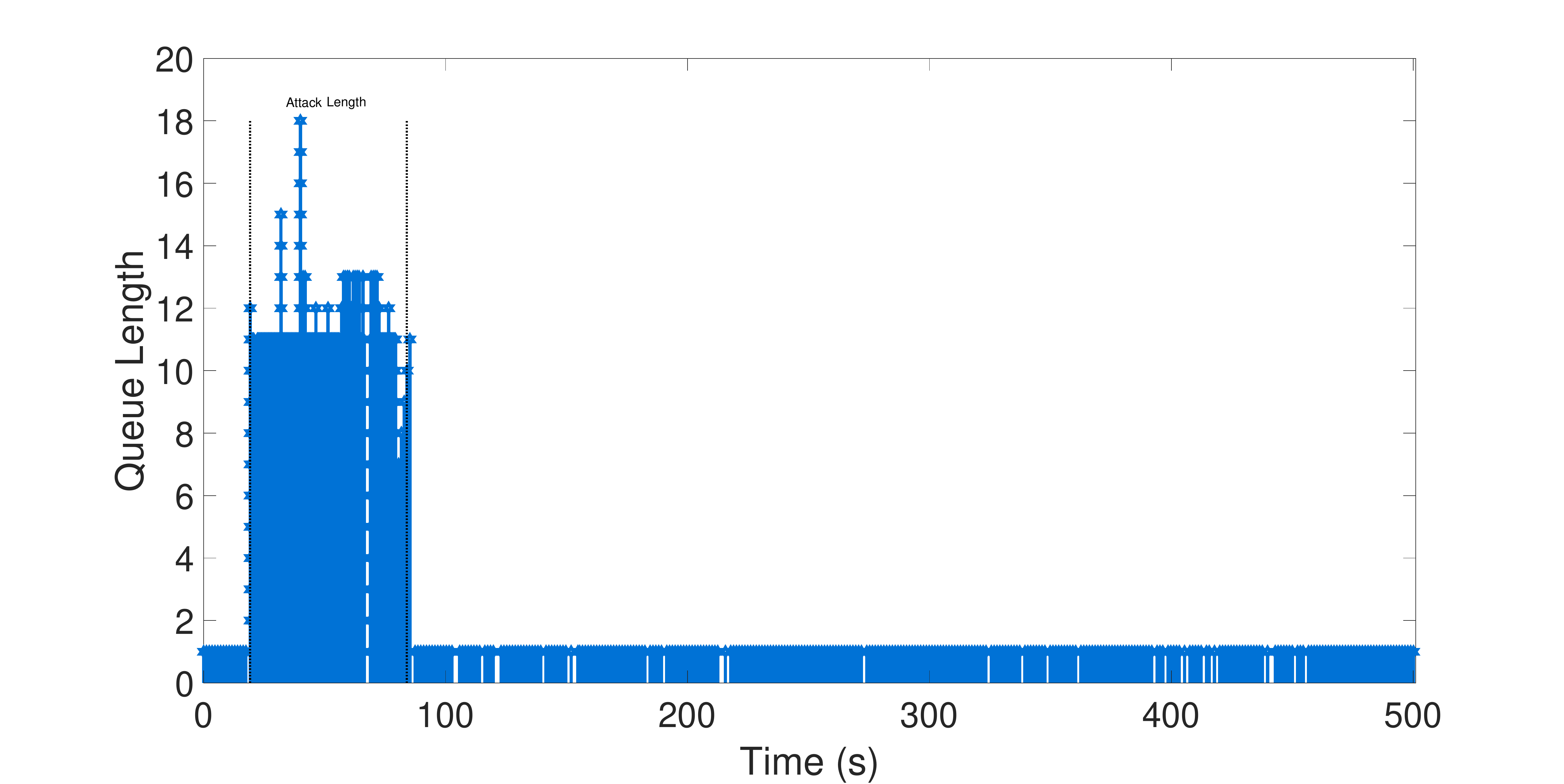}
	\caption{SQF queue length with $D=3~ms$ when the attack lasts for $60$ seconds, and the mitigation action is also applied.}
	\label{fig:QueueLengthMitigation60}
\end{figure}

Figure \ref{fig:QueueLengthMitigation60} displays the queue length of the SQF input buffer in a second experiment, when the attack lasts $60$ seconds, showing similar results to the first experiment. Both
measurements show the importance of having a simple mitigating action to deal with high volume Flood Attacks.

However, although this policy appears attractive, it comes at the cost of dropping legitimate (non-attack) packets that come from non-compromised IoT devices.

\section{Conclusions}

This paper has considered the effect of UDP Flood attacks on an IoT Server that processes incoming  traffic from
IoT devices via a local area network. The Server
also incorporates an AD module.  

We first show that such attacks, even when they last just a few seconds, create overload for the IoT Server, so that its normal operations, including AD, are substantially slowed down. 
We see, in particular, that a $60$ second attack may create a backlog of packets at the Server, that may require several hours to clear out. 

Thus we propose that the Server's input be ``protected''  by a special SQF front-end that operates the QDTP policy, in order to allow the timely operation of the Server even when an attack occurs. This approach requires that an inexpensive lightweight  hardware addition, such as an RPi, be installed
between the local area network that supports the IoT devices and the Server. Several experiments are used to illustrate the effectiveness of the proposed approach.

However, the SQF with its QDTP policy requires that a key timing parameter $D$ be chosen. Therefore, we provide a theoretical analysis of how $D$ should be selected: we show that it must be just larger than the AD processing time
of the Server under normal, i.e. non-attack, conditions. We then validate this observation with several experiments and show that the SQF can preserve the Server from congestion and overload, and allow it to operate normally. However, we note that the congestion that has been eliminated at the Server may now accumulate at the SQF input, although this in itself does not stop the RPi based SQF from continuing its normal operation. 

Furthermore, when the incoming traffic rate is such that it clearly indicates an attack, or when the Server informs the SQF that an attack is occuring, we can implement a mitigating action at the SQF to drop incoming packets in relatively short successive time intervals. This approach is tested experimentally and shown to be effective. However, the fact that such a policy may also drop incoming legitimate packets implies that there will be circumstances when it cannot be used and a close coupling between AD at the Server and packet drop actions at the SQF will be needed.

While this paper has focused on an architecture with multiple sources of IoT traffic represented by several RPis, future work will consider Edge Systems having multiple IoT servers, as well as multiple IoT devices and packet sources,  and will study the usage of dynamic policies for AD and traffic routing at the edge for complex IoT Server and SQF gateway architectures. 

Another important issue that should be addressed in future work is the energy consumption of such edge systems \cite{Frohlich}, so that dynamic management policies may be used to minimize energy consumption, as well as to optimize Quality of Service and Cybersecurity.

\section*{Acknowledgements}

The  support of the EU H2020 IoTAC Research and Innovation Action under GA No. 952684 is gratefully acknowledged.

\bibliographystyle{ieeetran}
\bibliography{references,references2,references3,references4,mybib,security_issues_references,references_arnn_conference}

\end{document}